# Four Layered Approach to Non-Functional Requirements Analysis

Gopichand Merugu  and Anandarao.Akepogu

Associate Professor of CSE, BVRIT
JNTUH, Hyderabad

Prof. of CSE and Principal, JNTUA Anantapur,
Andhra Pradesh, India

## Abstract

Identification of non-functional requirements is important for successful development and deployment of the software product. The acceptance of the software product by the customer depends on the non-functional requirements which are incorporated in the software. For this, we need to identify all the non-functional requirements required by all stakeholders. In the literature not many approaches are available for this purpose. Hence, we have proposed a four layered analysis approach for identification of non-functional requirements. The proposed layered approach has many advantages over non-layered approach. As part of this approach some rules are also proposed to be used in each layer. The approach is applied successfully on two case studies. The identified non-functional requirements are validated using a check list and in addition the completeness of the identified non-requirements is computed using a metric.

## 1. Introduction

Non-Functional requirements analysis is one of the most important activity in requirements engineering. The requirements for a system are the descriptions of the services provided by the system and its operational constraints. Non-functional requirements are the constraints over the services. The Functional requirements reflect the needs of customers for a system. The process of finding out, analyzing documenting and checking these services and constraints is called requirements engineering [1]. In requirements engineering requirements elicitation and analysis is an important activity. In this requirements elicitation and analysis we need to elicit and analyze both functional and non functional requirements. But unfortunately people are focusing more on functional requirements and giving low priority to nonfunctional requirements. It is always important to consider both equally in software development.

Though we satisfy the functional requirements if we cannot satisfy the nonfunctional requirements people may not accept the product. As software complexity grows and clients demand higher and higher quality software non-functional requirements can no longer be considered to be of secondary importance. Due to this we need to elicit and analyze the non functional requirements well.  There are few methods in literature for this purpose. But they are not that much effective and complete in elicitation and analysis. Each method has its own strengths and weaknesses. But if we use these approaches, it is probably difficult to understand the requirements of the system completely and correctly. This may cause to get failure during system development. Additionally, these methods do not support the non-functional requirements elicitation process well. For successful software development, we need the non functional requirements analysis method to be able to elicit and analyze the requirements of complex software system from all perspectives.  Hence, we propose an approach that facilitates the elicitation, analysis, and understanding of the various dimensions of requirements of complex software systems. This paper consists of five sections. Following the introduction, section 2 briefly describes the existing methods for requirements analysis and their strengths and weaknesses. The proposed Four Layered Non-functional requirements analysis approach is discussed in section 3. Section 4 discusses how the approach has been applied in developing Library management system and ATM System. Section 5 provides conclusion.

## 2. Related Work

A few non functional requirements (NFR) analysis methods have been proposed in the literature with tools supporting the analysis processes. Most of the early work on NFRs focused on measuring how much a software system is in accordance with the set of NFRs that it should satisfy, using some form of quantitative analysis [2] [14] [16] [24], offering predefined metrics to assess the degree to which a given software object meets a particular NFR.

A few number of works proposed to use approaches which explicitly deal with NFRs before metrics are applicable [7][3] [9][17]. These works propose the use of techniques to justify design decisions on the inclusion or exclusion of requirements which will impact on the software design. Unlike the metrics approaches, these latter approaches are concerned about making NFRs a relevant and important part of the software development process. Boehm and In [3] propose a knowledge base where NFRs are prioritized through stakeholders' perspectives, dealing with NFRs at a high level of abstraction. Kirner [16] describe properties for six NFRs from the real-time system domain: performance, reliability, safety, security, maintainability and usability. This work provides heuristics on how to apply the identified properties to meet the NFRs and later measure these NFRs. However, it lacks a broader approach that can be applied to other NFRs, in the real-time domain or in other domains. The NFR Framework is one of the few works to deal with NFR starting from the early stages of software development through a broader perspective. The NFR Framework [7]views NFRs as goals that might conflict among each other and must be represented as soft goals to be satisfied. The soft goal concept was introduced to cope with the abstract and informal nature of NFRs. Each soft goal will be decomposed into sub-goals represented by a graph structure inspired by the and/Or trees used in problem solving. This process continues until the requirements engineer considers the soft goal satisfied The other methods used widely are Scenario-based analysis[19] and Language extended lexicon(LEL)[25] . In Scenario based analysis , a method was proposed that describes Scenario templates for NFR with heuristics for Scenario generation elaboration and validation. The LEL is based on a code system composed of symbols where each symbol is an entry expressed in terms of notions and behavioral responses. The notions must try to elicit the meaning of the symbol and its fundamental relations with other entries. In the above said methods each method is having its own advantages and disadvantages. Since there is scope for a better method we propose an approach called "Four layered approach to non functional requirements analysis" which is very simple and easy to use.

Using the proposed approach we can identify NFRs from multiple views of stakeholders. We can save more of system development time compare to traditional approaches, and at the same time it supports agile software development.

# 3. Non-Functional Requirements Analysis Approach

As the complexity of software system increases non functional requirements elicitation and analysis are becoming increasingly difficult in software development. Non-Functional Requirements elicitation and analysis may involve a variety of people in an organization. The term stakeholder is used to refer to any person or group who will be affected by the system directly or indirectly. Single view forces us to look at the requirements only from a particular perspective. In order to elicit and analyze requirements completely multiple views needs to be considered to meet all stakeholder expectations. Even though different methods have been proposed each method has its own strengths and weaknesses. Hence we proposed a generalized approach based on stakeholders view. This is a Four Layered approach to non-functional requirements analysis. This approach includes some rules and non-functional requirements analysis process. Using this approach we can identify the goals, sub goals and finally non functional requirements. This is used to identify the all non functional requirements from multiple views of different stakeholder. The main objective of this approach is to find out non functional requirements which are very important to consider in any system.

The approach uses Four layered analysis. The four layers are shown in figure 1. The layered approach offers many advantages, and they are listed below.

- - Scalability : A layered approach scales better when compared to horizontal approach
- - Better Flexibility: Layered approach improves flexibility in terms of options and choices.
- - Cost Effective: The layered approach is the most economical way of developing and implementing any system. In this context developing a system means, identifying non-functional requirements for the system System development depends on how well we identify the requirements

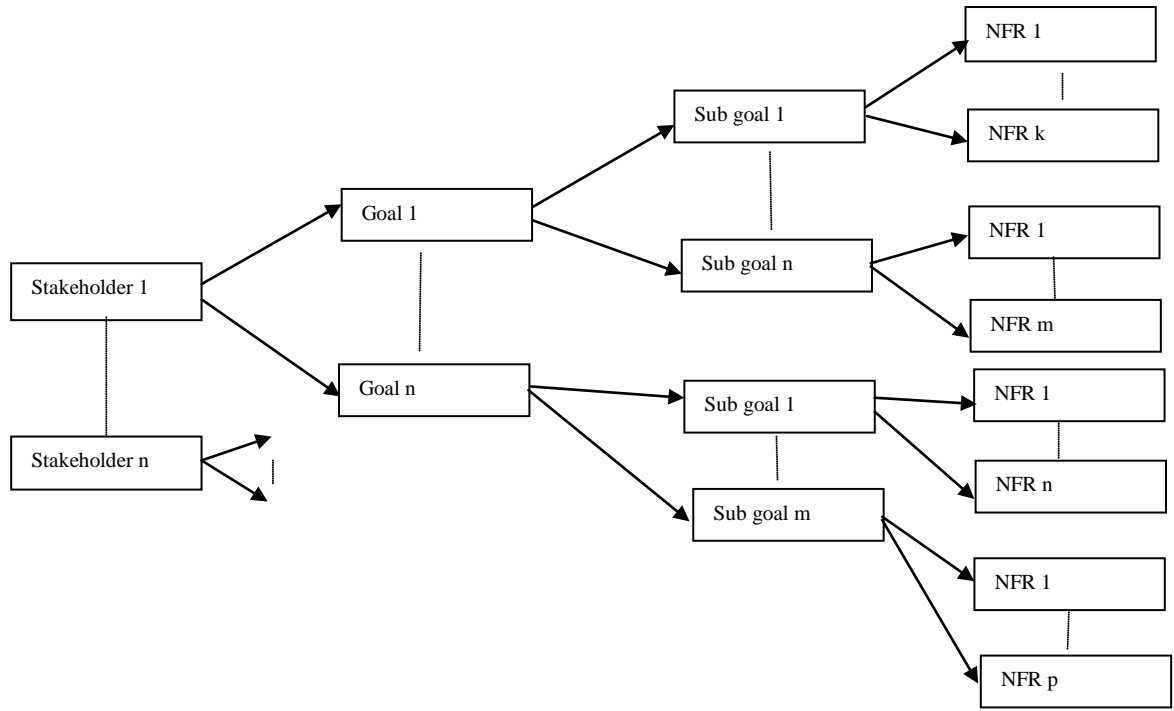

Fig. 1 General Architecture for Four layered analysis for non-functional requirements identification.

- Task segmentation: Breaking large complex systems into manageable subcomponents allows for easier development and implementation of any system.
- Enhanced Understanding: Layering allows for examination in isolation of subcomponents (processes/procedures) applicable for each layer and as well the whole.
- Rapid Application Development: Using the layered Approach requirements can be identified in less amount of time which leads fast application development.

Any software system requirements can be put broadly in the following two statements.

1. Each <system> provides <services> to <user>
   Each <service>must satisfy some <constraints> in order to meet customer needs.
2. Functional requirements state <what> the system supposed to do, and Non functional requirements state <how> the system supposed to be.

In this paper the mains focus is on non-functional requirements (constraints). The non functional requirements identification process can be divided into the four steps as follows and shown in activity diagram. (Fig. 2).

Step 1: Identify the key stakeholders of the system.
Step 2: Generate the goals from Stakeholders based on developers' Knowledge and experience.
Step 3: Decompose the goal into sub goals.
Step 4: Identify non-functional requirements for each sub goal.
The following rules are used in the above process.
  The rules are:
      <Who> are the stakeholders?
      <What> are the services (goals)
      <What > are the sub goals of each service?
      <How> the sub goals are achieved under constraints.
  The above five rules can be visualized in the following rule.
<Who> is stakeholder <What> are the services (goals) that system should provide to the stakeholder, <What> are the sub goals for the goals, <How> the system will perform sub goals under constraints
Key words like <who>, <what>, <how> are variables which can be assigned for specific words in the given context.
Key words used in the rules are:

Who: Stakeholders
What: Functional requirements/services
How: Non functional requirements/constraints

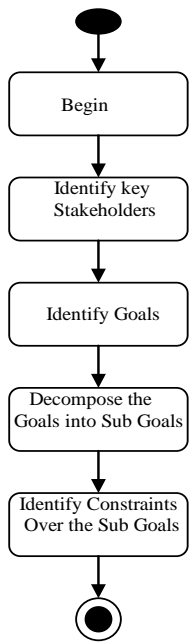

Fig. 2. Non-Functional requirements analysis process

## 3.1 Metrics for Non-Functional Requirements Specification

Metrics proposed by Davis[24] for identifying and measuring the quality in software requirements specification is used in this paper.

To determine completeness of overall requirements the metric used is:

$MCR = n_c / [n_c + n_{nv}]$

Where $n_c$ is number of requirements has been validated as correct and $n_{nv}$ is number of requirements that have not yet been validated. In this paper we have used this equation only for finding the completeness of the non-functional requirements

## 3.2 Non-Functional Requirements Validation

The following check list is used for validating non-functional requirements [23]:

- Does each requirement have source?
- Is each requirement achievable in the technical environment that will house the system?
- Each requirement is testable once implemented.
- Is each requirement bounded and unambiguous
- Do any requirements conflict with other requirements?
- Is the requirement traceable to goals of the system?
- Is the requirement is bounded in quantitative terms.
- Are requirements stated clearly? Can they be misinterpreted?

To validate our approach we have used the above mentioned check list which includes important questions.
We collected the information in the form of answers for these questions from various people both from Library System and ATM System about non-functional requirements.

## 3.3 Finding critical non-functional requirements.

The traceability matrix is used for finding critical non-functional requirements.

Table1. Traceability Matrix for Finding Critical Non-Functional Requirements

|      | G1 | G2 | G3 | G4 | Gm |
|------|----|----|----|----|----|
| NFR1 |    |    | x  |    |    |
| NFR2 |    | x  | x  |    | x  |
| NFR3 |    |    |    | x  |    |
| NFRn |    |    | x  |    |    |

In the above table the NFR1,NFR2,NFR3,NFRn represents non-functional requirements and G1 to Gm represents Goals. From the table we can say that NFR2 is critical, since many goals require NFR2.

# 4. CASE STUDY

In this section we have shown how the proposed approach helps to identify the non functional requirements from four layered requirements analysis. We considered two case studies these are online library system and ATM System. These systems are used by different stakeholders having different goals covering different perspectives. The proposed rules as part of this approach are used in identifying the non-functional requirements in two case studies.

## 4.1 Library Management System

The library system is used by many stakeholders like borrower, librarian and having their own goals and constraints. The constraints are identified using the rules.
Rule 1: <who> are the stakeholders?
The identified stakeholders are given in table 2.
Rule 2: <What> are the services (goals)
The identified services (goals) for all the stakeholders are given in table 2.
Rule 3: <What > are the sub goals of each service
The identified sub-goals for all the stakeholders are given in table 2, and sub goals for stakeholder "Member" are shown in figure 3.
Rule 4: <How> the sub goals are achieved under constraints.
The identified non-functional requirements for all the stakeholders are given in table 2, and non-functional requirements for stakeholders "Member" and "Librarian" are shown in figure 3.

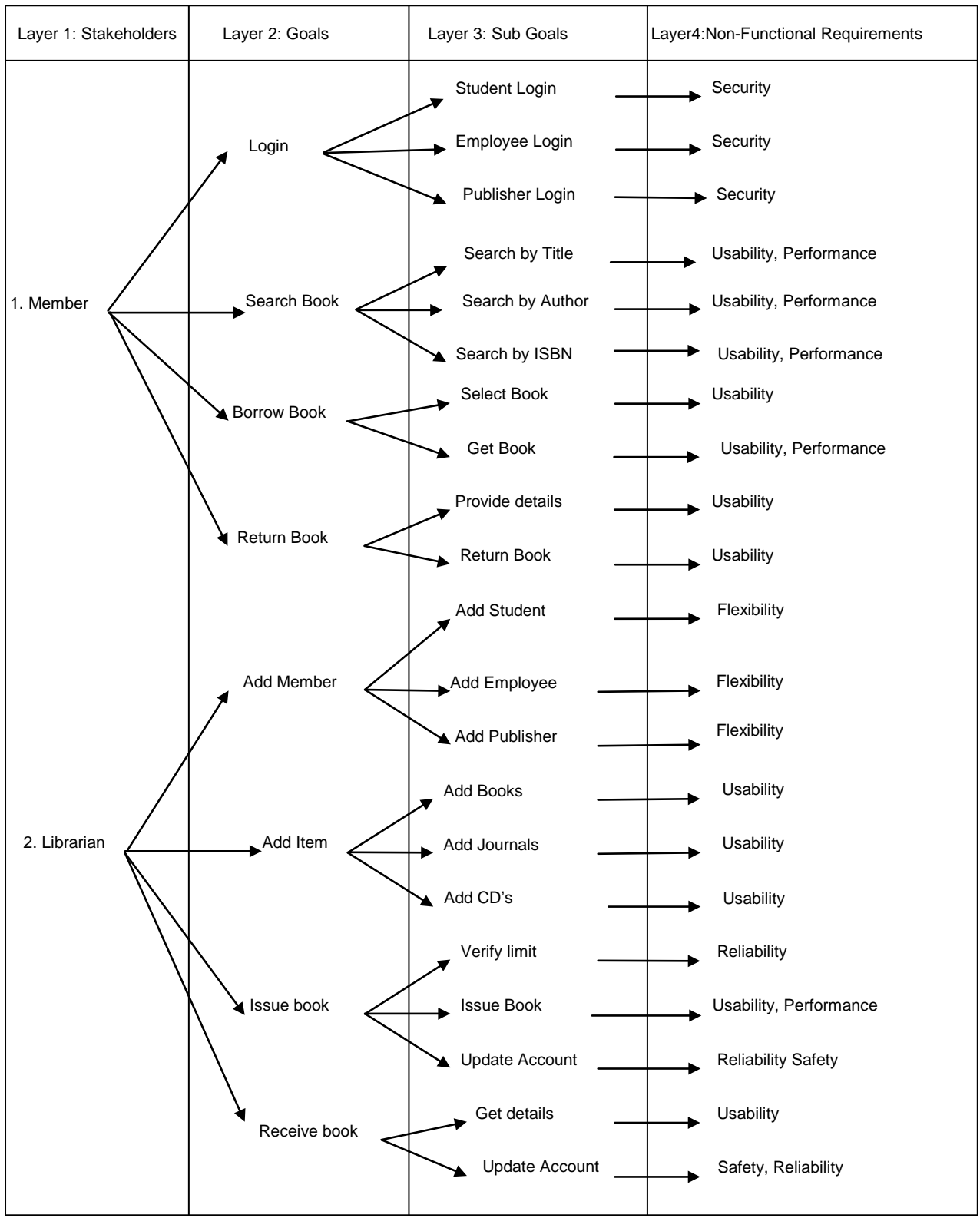

Fig.3 Four layered Requirements analysis sample for Library Management System

Table 2: Non-functional Requirements analysis sample for Library Management System

| System | Library management system |
|---|---|
| Stakeholders | Member, System Administrator, Librarian. |
| All Goals | Login, view catalog ,Search book, Reserve Book  Barrow book, Return book, pay fine Add item, update database, Register Member, Check Reports, Issue Book. |
| All Sub goals | Student login ,Employee login, Publisher login , view by subject, view by course, view by publisher , Search book by author , Search book by title ,Search book by ISBN, Request for book, Add book, Add journal, Add Cd's, Register student, Register Faculty, Register Publisher, Update book information, update barrower information, view the report, Edit report, Get book. |
| All Non functional requirements | Usability, Performance, Reliability, Security, Safety, Flexibility. |

The metric MCR is computed for Library System. The computed value of MCR for library system is :

$$MCR = n_c / [n_c + n_{nv}]$$

$$MCR = 6 / [6+0] = 1$$

The closer the value of MCR to 1 the maximum is the completeness of the requirements. In this paper we considered this equation for non-functional requirements only.

The requirements are validated against the checklist given in section 3.2. It is found that for all 8 questions the answer is yes. The validation metric is 8/8 = 1. Hence validated. Critical Non-Functional Requirements for Library System are identified using traceability matrix given in table.3.

Table 3: Traceability Matrix for Finding Critical Non-Functional Requirements for Library system

|  | G1 | G2 | G3 | G4 | G5 | G6 | G7 | G8 |
|---|---|---|---|---|---|---|---|---|
| U |  | X | X | X | X | X | X | X |
| P |  | X | X |  |  |  | X |  |
| $S_e$ | X |  |  |  |  |  |  |  |
| R |  |  |  |  |  |  | X | X |
| $S_a$ |  |  |  |  |  |  | X | X |
| F |  |  |  |  | X |  |  |  |

Here we made an effort for identifying critical non functional requirements. The above traceability matrix is for finding the critical non functional requirements for library system. The notations represented in rows U, P, $S_e$,R,$S_a$ and F are  Usability, Performance, Security, Reliability, Safety and Flexibility.  The corresponding columns represent goals like login, Search book, Borrow book, Return book, Add Member, Add Item, Issue Book,

Receive Book which are from Figure 3. So from the above table we can say that usability and performance are more critical than other non functional requirements. Similarly  we can find The critical NFRs which are identified must be implemented in the system. The acceptance of the system by the user is more likely if critical NFRs are implemented in the system.

## 4.2 ATM System

In   ATM system main stakeholders are like customer, system and administrator. Each stakeholder will have their own goals and constraints. The constraints are identified by using the rules.

Rule 1: <who> are the stakeholders?
The identified stakeholders are given in table 3.
 Rule 2: <What> are the services (goals)
The identified services (goals) for all the stakeholders are given in table 3.
Rule 3: <What > are the sub goals of each service
The identified sub-goals for all the stakeholders are given in table 3, and sub goals for stakeholder "Customer" are shown in figure 4.
Rule 4: <How> the sub goals are achieved under constraints.
The identified non-functional requirements for all the stakeholders are given in table 2, and non-functional requirements for stakeholder "Customer" are shown in figure 4.
The constraints (NFR) identified for the stakeholder "Member" are given in figure 3 and all the NFRs for all stakeholders are given in table 3. The layered analysis for stakeholder "Customer'' in first layer is given in figure 4.

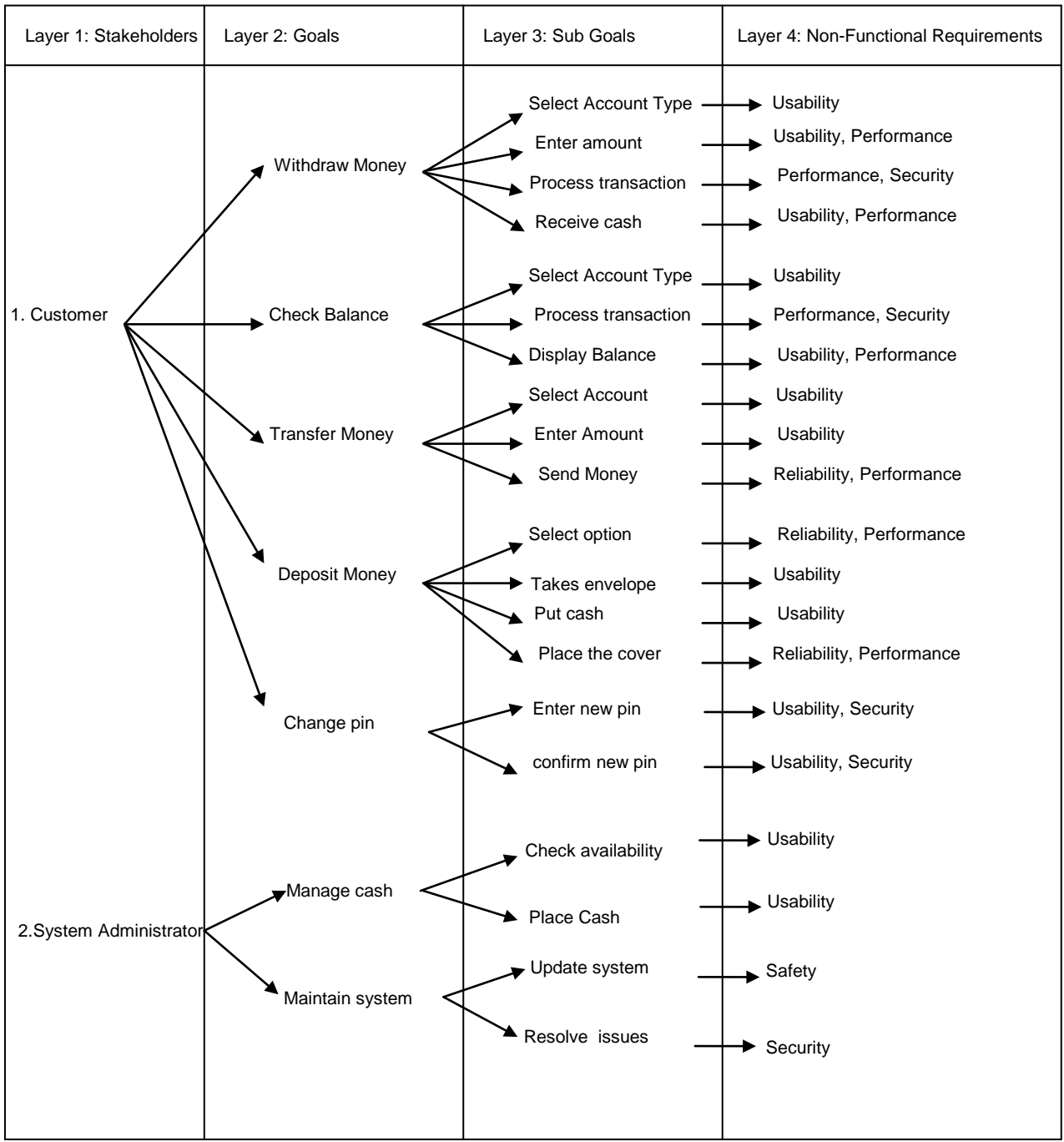

Fig 4: Four layered requirements analysis sample for ATM System

Table 4: Requirements analysis sample for ATM System

| System | ATM System |
|---|---|
| Stakeholders | Customer, System Administrator |
| All Goals | Withdraw money, Deposit money, Transfer money, Check Balance, Get mini statement, Change password, Verify card, Display Information, Process Transaction, Update Account, Update Database, Print Receipt. |
| All Sub goals | Enter Pin number, Receive Pin number, Verify Pin Number, Send result, Select account type, enter amount, Check account type, Process transaction, Dispatch cash, Receive cash, Select option, takes envelope, Put cash, Place the envelope, Select account, enter amount, send money, Display Account Information, Receive details, Update Details. |
| All Non-functional requirements | Usability, Performance, Reliability, Security, Safety. |

The metrics MCR is computed for ATM System. The computed value of MCR for ATM system is :

$MCR = n_c / [n_c + n_{nv}]$

$MCR = 5 / [5+0] = 1$

The closer is the value of MCR to 1 the maximum is the completeness of the requirements.

The requirements are validated against the checklist given in section3.2. It is found that for all 8 questions the answer is yes. The validation metric is the 8/8 = 1. Hence identified NFRs are validated.

Table 5: Traceability Matrix for Finding Critical
Non-Functional Requirements for ATM system

| | G1 | G2 | G3 | G4 | G5 | G6 | G7 |
|---|---|---|---|---|---|---|---|
| U | X | X | X | X | X | X | |
| P | X | X | X | X | | | |
| $S_e$ | X | X | | | X | | X |
| R | | | X | X | | | |
| $S_a$ | | | | | | | X |

Here we made an effort for identifying critical non functional requirements. The above traceability matrix is for finding the critical non functional requirements for ATM system. The notations represented in rows U, P, $S_e$, R and $S_a$ are Usability, Performance, Security, Reliability and Safety.

The corresponding columns represent Goals like With draw money, Check balance, Transfer money and Deposit Money these are from figure 4. So from the above table we can say that usability Performance and Security are more critical than other requirements.

The critical NFRs which are identified must be implemented in the system. The acceptance of the system by the user is more likely if critical NFRs are implemented in the system

## 5. CONCLUSIONS

The acceptance of any software product by the customer depends on how well we identify the Non-Functional requirements and incorporates them in the software. In this paper we have proposed an approach for identifying Non-functional requirements. The approach is based on four layered analysis. By this approach we can identify all the Non-Functional requirements required by all stakeholders. As part of this approach we have proposed some rules to be used in the identification process, metrics for non-functional requirements completeness and Check list for requirements validation. The approach is applied successfully on two case studies.

The advantages of the approach are:

1. No analysis modeling is required so that we save can save time.

2. Since non functional requirements are found, the Probability of user acceptance of software is more.

3. All the goals are found by considering multiple views hence no loss of information. By applying this approach we can find all possible non functional requirements.

4. Functional requirements along with non functional requirements acts as validation criteria to be included in SRS

5. This approach can be used in the context of all process models including agile process model.

**Prof. Ananda Rao Akepogu** received B.Sc. (M.P.C) degree from Silver Jubilee Govt. College, SV Univer-sity, Andhra Pradesh, India. He received B.Tech. degree in Computer Science & Engineering and M.Tech. degree in A.I & Robotics from University of Hyderabad, Andhra Pradesh, India. He received Ph.D. from Indian Institute of Technology, Madras, India. He is Professor of Computer Science & Engineering and Principal of JNTU College of Engineering, Anantapur, India. Prof. Ananda Rao published more than fifty research papers in international journals, conferences and authored three books. His main research interest includes software engineering and data mining.

**GopiChand.Merugu** is pursuing Ph.D. in Computer Science & Engineering from JNTUA, Anantapur, India and he received his M.Tech. in Computer Science & Engineering from the same university. He received B.Tech. degree in Information Science & Technology from Nagarjuna University, India.He is Associate Professor of computer science, BVRIT, JNTUH, Hyderabad. He is a member of IEEE,CSI and IAENG.